# Momentum-space imaging of ultra-thin electron liquids in δ-doped silicon

Procopios Constantinou[*,1,2,5], Taylor Stock[1,9], Eleanor Crane[1,9], Alexander Kölker[1,9], Marcel van Loon[1,2], Juerong Li[3], Sarah Fearn[1,4], Henric Bornemann[1,2], Nicolò D'Anna[5], Andrew Fisher[1,2], Vladimir N. Strocov[5], Gabriel Aeppli[5,6,7,8], Neil Curson[1,9], Steven Schofield[†1,2]

[1] *London Centre for Nanotechnology, University College London, WC1H 0AH, London, UK*
[2] *Department of Physics and Astronomy, University College London, WC1E 6BT, London, UK*
[3] *Advanced Technology Institute, University of Surrey, Guildford GU2 7XH, UK*
[4] *Department of Materials, Imperial College of London, London SW7 2AZ, UK*
[5] *Paul Scherrer Institute, 5232 Villigen, Switzerland*
[6] *Institute of Physics, Ecole Polytechnique Fédérale de Lausanne (EPFL), 1015 Lausanne*
[7] *Department of Physics, ETH Zürich, 8093 Zürich*
[8] *Quantum Center, Eidgenössische Technische Hochschule Zurich (ETHZ), 8093 Zurich, Switzerland*
[9] *Department of Electronic and Electrical Engineering, University College London, London WC1E 7JE, UK*

[*] *procopios.constantinou@psi.ch*
[†] *s.schofield@ucl.ac.uk*

(Dated: Friday, 29 September 2023)

**Abstract:** Two-dimensional dopant layers (δ-layers) in semiconductors provide the high-mobility electron liquids (2DELs) needed for nanoscale quantum-electronic devices. Key parameters such as carrier densities, effective masses, and confinement thicknesses for 2DELs have traditionally been extracted from quantum magnetotransport. In principle, the parameters are immediately readable from the one-electron spectral function that can be measured by angle-resolved photoemission (ARPES). Here, we measure buried 2DEL δ-layers in silicon with soft X-ray (SX) ARPES to obtain detailed information about their filled conduction bands and extract device-relevant properties. We take advantage of the larger probing depth and photon energy range of SX-ARPES relative to vacuum ultraviolet (VUV) ARPES to accurately measure the δ-layer electronic confinement. Our measurements are made on ambient-exposed samples and yield extremely thin (< 1 nm) and dense (~$10^{14}$ cm$^{-2}$) 2DELs. Critically, we use this method to show that δ-layers of arsenic exhibit better electronic confinement than δ-layers of phosphorus fabricated under identical conditions.

Two-dimensional (2D) quantum-confined electronic systems have long been venues for discoveries in fundamental physics and the development of new devices [1]. Technological 2D systems have traditionally consisted of planar heterostructures and field-effect devices, particularly in compound semiconductors [2]. In recent years, there has similarly emerged strong interest in 2D electron states in van der Waals systems, such as graphene, and the transition metal dichalcogenides for future nanoscale and quantum-electronic devices [3]–[5]. Understandably, there is also strong interest in fabricating 2D electron states in the world's leading technological semiconductor, silicon. This is largely driven by the requirements of proposed nano- and quantum-electronic applications employing atomically abrupt dopant profiles, e.g., the famed Kane solid-state quantum computer and related designs [6]–[8]. 2D electron states can be created in silicon via so-called δ-doping, which involves the physical [9] or chemical [10] deposition of dopant atoms onto a silicon surface, followed by silicon overgrowth to produce sharp, 2D doped layers (Figure 1a). At high doping concentrations, such δ-layers yield quantum-confined 2D conductive planes with electronic properties significantly different to those of the bulk silicon host [11].

The thinnest δ-layers prepared in silicon to date have relied on the chemical delivery of phosphorous [10], arsenic [12] or boron [13], with the resulting out-of-plane atomic distributions of dopant atoms having ~1 nm thicknesses [14]–[17]. The electronic thicknesses of these layers have also been estimated using quantum magnetoresistance [18], with similar results [19]. Such thicknesses are comparable to the wavelength of the conduction electrons, and the corresponding energy level quantisation was observed in planar junction tunnelling spectroscopy more than three decades ago [9], [20], [21]. Vacuum ultraviolet angle-resolved photoemission spectroscopy (VUV-ARPES) measurements of phosphorous δ-layers in silicon have also revealed quantised states, yet the origin of these quantised states was incorrectly attributed to the more exotic degeneracy lifting mechanism, valley interference [22]–[25]. To justify the anomalously large valley splitting energies reported, the authors cited density functional theory (DFT) calculations that were made for perfectly ideal, one-atom-thick δ-layers. However, DFT calculations of δ-layers with even a single atom deviation from a perfectly-thin δ-layer show the valley splitting reduces to ~1 meV [26]. Such small valley-splitting energies cannot presently be observed in ARPES measurements, and it has since been acknowledged that the observed splitting is due to confinement [27], [28], as first suggested in the 1980s [9], [20], [21]. Moreover, as discussed in Refs. [22], [23], the short inelastic mean free path of the ejected electrons in VUV-ARPES ($\lambda_e \approx 0.5$ nm) means the signal for previous ARPES measurements [23], [28], [29] does not directly originate from the δ-layer (that is up to $4\lambda_e$ beneath the surface), but is instead a near-surface resonance enhancement that enables only a small fraction of the wavefunction to be probed [23]. Furthermore, because VUV-ARPES has limited momentum resolution along the surface normal, it was impossible to measure a corresponding momentum spread whose inverse would be the key parameter of the 2DEL, namely the electronic thickness, from which the origin and level quantisation of the 2DEL can be deduced.

In this paper, we report comprehensive soft X-ray ARPES (SX-ARPES) measurements of δ-layers in silicon. The high photon energies of SX-ARPES ($hv$ = 300 - 1600 eV) give access to a much longer electron mean free path ($\lambda_e \approx 2$ nm), which permits the extraction of electrons from depths of several nanometres beneath the surface [30]. This enables us to directly probe δ-layers underneath the native surface oxide of samples exposed to ambient after their fabrication, whilst maintaining a very sharp out-of-plane $k_z$ momentum resolution, $\Delta k_z$, which is equal to $\Delta k_z = \lambda_e^{-1}$ [31]. Our experiments therefore differ qualitatively from the previous VUV-ARPES [22]–[25]. We present, for the first time, energy and momentum maps resolved with high momentum resolution in the plane perpendicular to the δ-layer, revealing the detailed δ-layer band structure in the $k_z$-$k_\parallel$ plane. Our measurements conclusively demonstrate that the δ-layer band structure is non-dispersive in the plane perpendicular to the δ-layer in a manner significantly more convincing than a previous attempt using VUV-ARPES $k_z$-binding energy scans [22]. Moreover, exactly as for photoemission tomography of molecules [32]–[34], our $k_z$ momentum dependencies are related via a Fourier transform to electron densities in real space, and thus measure directly the real-space thicknesses of the occupied quantised electronic states that constitute the 2DEL. We apply this method to investigate the optimisation of δ-layer electronic thickness in silicon, and to compare δ-layers fabricated with arsenic and phosphorus. We show that arsenic δ-layers are significantly more electronically confined than phosphorus δ-layers prepared under identical conditions, and we determine the carrier density via a Luttinger analysis of the Fermi surface.

Our SX-ARPES experiments feature an X-ray spot size of (10 × 73) μm$^2$, which is comparable to the size of the Hall-bars used for quantum magnetotransport measurements. Next-generation light sources together with new optics will enable SX-nanoARPES with better energy resolution and sub-micron spot sizes [35], thus providing a tool complementary to X-ray inspection of integrated circuit morphology [36], [37] and chemical composition in the sense that it will image the electrons switched in devices. While such ARPES measurements have already been conducted in the UV regime [38], [39], extension to the SX regime will offer an enhanced bulk sensitivity for probing buried heterostructures or interfaces. Although scanning microwave microscopy [40] also images the conduction electrons in devices, it does not yield their three-dimensional momentum distribution. However, SX-nanoARPES, along with the methods and analysis we present here, can do so, greatly expanding the possibilities for characterizing semiconductor nanostructures and devices.

**Background**

The dynamic behaviour of conduction electrons in bulk silicon is determined by a set of 6 degenerate conduction band valleys, with minima at equivalent points in reciprocal space along the <100> directions [41]. Bulk electron doping causes these valleys to become occupied and, at high doping levels, will result in ellipsoidal Fermi surfaces, one around each minimum (Figure 1b). However, when electrons are confined to 2D planes, as for δ-doping, the Bloch wavevector component in the $k_z$ direction is no longer a good quantum number, and the energy becomes quantised into discrete levels, $E_n$. The in-plane wavevector components $k_x$ and $k_y$ remain good quantum numbers and the electronic states can be described using the formalism of effective mass theory [42].

According to elementary quantum mechanics, the degree of confinement is governed by the potential created by the δ-layer, the effective mass of the electrons, and the number of wavefunction nodes. Since the δ-doping breaks the degeneracy of the six valleys, the two valleys centred at $k_x = k_y = 0$ are characterised by a single, in-plane, transverse effective mass and the quantised states are correspondingly labelled $n\Gamma$ (where $n$ is the subband number), while the remaining four in-plane valleys are characterised by in-plane longitudinal and transverse effective masses and are labelled $m\Delta$ (where $m$ is the subband number) [43]–[45]. Subsequently, in the direction of quantisation the $n\Gamma$ and $m\Delta$ subbands derive from bands with a heavy and light effective mass respectively, leading to different spectra for states derived from different valleys. The right-hand panel of Figure 1a shows a self-consistent Schrödinger-Poisson model of how the $n = 1$ and $n = 2$ wavefunctions (labelled 1Γ and 2Γ) for electrons with a heavy mass bracket the $m = 1$ wavefunction (labelled 1Δ) for the lighter, and hence less confined, electron; the simulation in Figure 1a was performed using the electron density and electronic thickness extracted from our SX-ARPES measurements of a 2 nm overgrown arsenic δ-layer, as described below. Moreover, our calculations treat the $n\Gamma$ and $m\Delta$ subbands as standing wave solutions that originate from the superposition of two plane waves moving with $\pm k_z$ momenta, confined by the boundary of the δ-layer and in the absence of so-called valley interference [11].

In practice, the δ-layer wave function is characterised by an envelope function in the z-direction that decays with distance away from the δ-layer, combined with an oscillatory Bloch wave component established by the bulk conduction states from which the δ-layer is derived. The Fourier spectrum of such a state is peaked about the values of $k_z$ corresponding to its Bloch wave origins and is oscillatory in $k_z$ at multiples of the reciprocal lattice vector [30], [46], [47]. Thus, the Fermi surface picture of Figure 1b is transformed by the replacement of conduction ellipsoids with states that do not disperse in $k_z$, and can be visualised, from the standpoint of an ARPES experiment, as being cylindrical or elliptic-cylindrical in shape (Figure 1c); the extent of these states in $k_z$ is inversely proportional to the electronic (not chemical) real space thickness of the δ-layer [25], [30]. A 2D system confined along z by an infinitely deep and infinitesimally narrow potential would yield states with infinitely long profiles along $k_z$, while at the other extreme, for a fully three-dimensional doped system, the states should return to reside within the ellipsoidal Fermi-surfaces shown in Figure 1b. For real layers of some finite thickness, a phenomenological equation for the thickness of the layer is [30]:

$$\delta z = \frac{1}{\delta k_z - \delta k_\infty}, \qquad (1)$$

where $\delta k_z$ is the extent of the 2D valley state in $k_z$, and $\delta k_\infty$ is the corresponding length of the state for the same electron doping level in the absence of 2D confinement. We determine $\delta k_z$ and $\delta k_\infty$ experimentally from our SX-ARPES data by measuring the longitudinal extent of the out-of-plane (Γ) valley, and the in-plane (Δ) valleys respectively. Careful measurement of these quantities and application of Equation 1 thus produces a direct measure of the electronic thickness, $\delta z$, of the δ-layers.

Figure 1d summarises our results for the electronic thickness of the δ-layer. Here we show the longitudinal extent of the in-plane and out-of-plane valleys versus their transverse extent. The data clusters into two groups, for the Γ and Δ valleys, respectively. In particular, the Δ valleys lie along a straight line characterising the ellipsoidal shape of the bulk silicon conduction band valleys (as set by the ratio of the bulk longitudinal and transverse effective masses). In stark contrast, the Γ valleys appear elongated in the longitudinal direction and are therefore grouped together in the top left of the plot. This lengthening of the states in $k_z$ is characteristic of 2D electronic states, to be discussed further below.

**δ-layer carrier density and Fermi-surface measurements**

We fabricated δ-layer samples using either phosphorus or arsenic as the dopant species. The *Methods* sections gives details of the sample preparations. Secondary ion mass spectrometry (SIMS) and Hall effect measurements confirmed the anticipated highly peaked dopant distributions and dopant electrical activations for all samples (see *Supplementary Information*).

In Figure 2 we show the SX-ARPES Fermi surface maps acquired from a phosphorus (Figure 2a-d) and an arsenic (Figure 2e-h) δ-layer. The schematic Brillouin zone diagrams at the left of the figure illustrate the planes through which each of the Fermi surface slices have been taken: Figure 2b,f show $k_x$-$k_z$ slices that cut through two Γ and two Δ valleys, illustrated by the purple plane in the schematics. Figure 2c,g and Figure 2d,h show $k_x$-$k_y$ slices at different $k_z$ values, as indicated by the green and orange planes in the schematics, respectively.

The degeneracy breaking due to δ-layer confinement is readily apparent for both samples: the four Δ-valleys in the $k_x$-$k_y$ slices (Figure 2c,g) are uniform in size and shape, as expected, while in the $k_x$-$k_z$ slices (Figure 2b,f) we find the two Γ-valleys (at $\pm k_z$) appear significantly larger and brighter than the Δ-valleys. The main difference in intensity occurs because of the different in-plane effective masses of the two types of valleys, resulting in a different electronic density of states and hence measured spectral weights [44].

We can determine the 2D carrier density of the samples by analysing the area enclosed by each valley in the $k_x$-$k_y$ plane; in other words, determining the total area enclosed by the four Δ valleys in Figure 2c,g and also the $k_x$-$k_y$ slice through the two Γ valleys, one of which is shown in Figure 2d,h. We find that the resulting total carrier density for all samples lie within the range $(0.88 \pm 0.10) \times 10^{14}$ cm$^{-2}$, consistent with Hall effect measurements for all but one of the samples considered (see *Supplementary Information*). This concurs with our expectations, as at the self-saturation limit of δ-doping, 1 in every 4 silicon (001) surface atoms is replaced with a dopant, corresponding to a density of $\approx 1.4 \times 10^{14}$ cm$^{-2}$ [48]. We attribute the reduced measured carrier density to the deactivation of some donors via effects such as clustering (particularly for arsenic) [49], [50] and chemical interaction with oxygen atoms where the native oxidation of the surface and δ-layer overlap. Furthermore, we find that the carriers are equally distributed within the Γ and Δ subbands (see *Supplementary Information*), in agreement with the theoretical predictions of Ref. [42] and our own Schrödinger-Poisson modelling (Figure 1a), in contrast to previous VUV-ARPES that showed an unoccupied Δ band [27].

**δ-layer thickness determination**

As discussed above, an electronically 2D δ-layer should be dispersionless in $k_z$, and therefore its Γ valley should be a regular cylinder, rather than ellipsoidal. In addition, the extent of the state in $k_z$ provides a direct measure of the confinement thickness of the state. With this in mind, we have performed a quantitative analysis of four δ-layer samples, as shown in Figure 3. Two of the samples were phosphorous δ-layers and two were arsenic δ-layers, and for each dopant species we have performed a nominal silicon overgrowth of 2 nm and 3 nm. Figure 3a summarises our approach to determine the δ-layer confinement from the high-resolution Fermi surface maps of the $+k_z$ Γ-valleys (Figure 3d-g), and a comparable $+k_y$ Δ-valley (Figure 3b). We note that measurements were also made on samples overgrown with 1 and 4 nm of silicon. For the former, no conduction states were observed, which we attribute to the complete oxidation of the δ-layer when the sample was exposed to ambient for transport to the synchrotron. For the latter, the spectral intensity of the conduction states became incredibly weak, due to the electron escape depth being smaller than the δ-layer depth, making the analysis extremely difficult.

We have used an automated procedure to extract the edges of the $+k_z$ valleys: for each horizontal line-profile cut of the Fermi surface, we find the edges of the valleys, whose positions are shown as pairs of white dots on Figure 3d-g. For the arsenic δ-layer samples, two distinct peaks in each line-cut along $k_x$ are resolved and tracked. These two peaks correspond to the cusps of the parabolic dispersion of the electrons in $k_x$. For the phosphorous δ-layer samples, the peaks along $k_x$ could not be resolved directly, so instead the FWHM was measured. For each value in $k_z$, the separation between these two dots along the $k_x$ direction gives a measure of the Fermi wavevector, $k_F$, and these values of $k_F$ are plotted against $k_z$ in the corresponding panels Figure 3h-k. For each of the four δ-layer samples, we see that $k_F$ remains constant as a function of $k_z$ to within the uncertainties of our measurements, demonstrating that each of the four samples are dispersionless in $k_z$, as expected. For comparison, in Figure 3b,c, we apply the same analysis to one of the in-plane Δ valleys to plot $k_F$ as a function of $k_y$. Here we see that $k_F$ is not constant, but instead exhibits the expected dispersion corresponding to the longitudinal effective mass, from which we extract a value of $(0.90 \pm 0.05)m_e$, in agreement with its accepted value [51].

The analysis in Figure 3h-k provides a measure of the length of these features in $k_z$, i.e., $\delta k_z$. We obtain the corresponding 3D width, $\delta k_\infty$ from the analysis of the in-plane valley in Figure 3c. Using these values, we then extract the real space electronic thickness of the δ-layer using Equation 1. We find that for the arsenic δ-layer samples, $\delta z$ = 5.4 ± 0.1 Å, whereas for the phosphorus δ-layer samples, $\delta z$ = 9.7 ± 4.1 Å. A summary of the δ-layer thickness measurements using SIMS and SX-ARPES is shown in Table 1, where the physical dopant confinement and electronic thicknesses are stated respectively. In all cases, we find that arsenic δ-layers offer a better confinement relative to phosphorus, achieving sub-nm electronic thicknesses. We attribute this to the smaller diffusion coefficient of arsenic in silicon [52], which, under the same preparation conditions, sustains a more confined δ-layer than phosphorous [12]. Additionally, the δ-layer thickness was further confirmed by directly fitting the ARPES $k_z$-response to the convolution of Lorentzian spectral functions and by taking the Fourier Transform of the probability density function solutions from a Schrödinger-Poisson model of δ-layers (see *Supplementary Information*). In all instances, a mutual agreement was found.

### δ-layer subband energies and comparing to theory

The analysis of Figure 2 and Figure 3 provide, for each of our samples, a measure of the carrier density and electronic thickness, respectively. These parameters can be used to create an electrostatic model of the δ-layer (Figure 1a right) that we have used as the basis of self-consistent Schrödinger-Poisson modelling of the state quantisation in $k_z$ (details of the calculations can be found in *Supplementary Information*). Based on these measured parameters, our calculations show that each of our δ-layer samples should support 1Γ, 2Γ and 1Δ states. Additionally, in good agreement with our results, Figure 4b shows that the occupancy of the δ-layer subbands is also distributed evenly amongst the valleys, in good agreement with our experimental results [42].

To further compare these calculations with experiment, we have measured the in-plane band dispersion and $k_z$ state quantization directly. Figure 4c-f show measurements of the band dispersion, $E_B(k_x)$, taken through the centroid of the $+k_z$ valley for each of the four samples discussed in Figure 3. We have performed a careful two-component fit to this data [23], analysing both iso-$E_B$ and iso-$k_x$ slices for each data point, as illustrated on the side and top of each panel in Figure 4c-f. Each dataset is best described by two parabolic dispersions, readily interpretable as the 1Γ and 2Γ states expected from the theoretical calculations. A similar analysis of the Δ valley dispersion is provided in the *Supplementary Information*, showing in this case that only a single 1Δ state is observed experimentally. The measured binding energies of these states have been added to the theoretically predicted curves in Figure 4a, and there is good agreement between our calculated and measured band energies in each case.

**Conclusions**

We have presented the most comprehensive SX-ARPES measurements of dopant δ-layers in silicon, and revealed that at the high arsenic densities considered, there are three flavours of electrons derived from their confinement along the transverse and longitudinal directions of the conduction band minima of bulk silicon. Our data show that the arsenic δ-layer samples host the thinnest technological 2D electron liquids ever fabricated in silicon and are close to ideal 2D electron systems with a thickness comparable to the silicon lattice parameter; our thinnest arsenic δ-layer has an electronic thickness of 0.45 ± 0.04 nm. Moreover, we compared arsenic and phosphorus δ-layer samples and found that in all cases, the arsenic samples outperformed the phosphorus ones in two-dimensionality. All our samples are technologically relevant, having been exposed to ambient after their fabrication, demonstrating the remarkable stability of these ultra-thin, dense δ-layer systems and the capability of SX-ARPES to fully characterise their conduction bands directly and non-destructively. The fact that we can engineer such ultrathin high carrier density liquids represents yet another capability which can be exploited for new nano- and quantum-electronic applications in silicon.

**Methods**

**Sample fabrication:** Silicon $n$-type (10 Ω cm) Si(001) substrates were degassed and flash annealed to ~ 1200°C under ultra-high vacuum ($< 5 \times 10^{-10}$ mbar). This procedure is known to produce atomically clean surfaces with uniform atomically flat terraces of with widths of 10s to 100s of nanometres [53]. The atomically clean and flat surfaces were exposed to a saturation dose of phosphine or arsine, and then annealed at 350°C for 2 minutes to substitutionally incorporate the dopants. The dopant layer was then encapsulated by overgrowing either 2 or 3 nm of silicon using a silicon sublimation source, with a deposition rate of 1 ML/min. During the silicon overgrowth, we controlled the temperature of the sample in three steps to maximise the dopant confinement, following the so-called locking-layer procedure [12], [15]: the first 1.3 nm of silicon was grown at room temperature, followed by a rapid thermal anneal at 500°C for 15 s and a low-temperature epitaxial growth at 250°C for the remainder of the overgrowth. The samples were then removed from vacuum and exposed to ambient for their transport to the soft X-ray ARPES facility [54] at the Swiss Light Source.

**SX-ARPES experiments:** The ARPES measurements were performed at the soft X-ray ARPES facility [54] of the ADRESS beamline [55] at the Swiss Light Source, PSI, Switzerland. The accessible photon energy range is $h\nu$ = 300 – 1600 eV, with a photon flux of up to $10^{13}$ photons / s / (0.01% BW). To maximise the coherent spectral function (impaired by the thermal atomic motion [56]), the experiments were performed at a base temperature of 12 K, using circular polarised light. The combined (beamline and analyser) energy resolution varied from 50 meV at $h\nu = 400$ eV, to 90 meV at around 700 eV. The photoelectron momentum $k_x$ was directly measured through the emission angle along the analyser slit, $k_y$ is varied through the tilt rotation and $k_z$ is varied through $h\nu$. The angular resolution of the ARPES analyser (PHOIBOS-150) is 0.1°. Other relevant details of the SX-ARPES experiments, including experimental geometry can be found in Ref. [54].

**Acknowledgements**

We acknowledge helpful discussions with Oliver Warschkow, the beamtime support provided by Alla Chikina and Niels B. M. Schröter and the excellent technical support from Leonard Nue. The project was financially supported by the Engineering and Physical Sciences Research Council (EPSRC) project EP/M009564/1, the EPSRC Centre for Doctoral Training in Advanced Characterisation of Materials (EP/L015277/1), the Paul Scherrer Institute (PSI) and the European Union Horizon 2020 Research and Innovation Program, within the Hidden, Entangled and Resonating Order (HERO) project (810451). Procopios Constantinou was partially supported by Microsoft Corporation.

**Data availability statement**

The data that support the findings of this study are openly available on Zenodo ([zenodo.org](zenodo.org)) at https://doi.org/10.5281/zenodo.7813819.

**Table 1: Quantifying the δ-layer confinement.** Two independent measures of the δ-layer confinement using secondary ion mass spectrometry (SIMS) and soft x-ray ARPES (SX-ARPES) experiments. All SIMS profiles are shown in the *Supplementary Information* and we note a general agreement with prior measurements in [12], [14]–[17]. For the investigated samples, arsenic δ-layers consistently yield a better confinement relative to phosphorous. Note that SIMS measures an upper bound on the physical thickness of the δ-layer dopant distribution, whereas SX-ARPES measures the electronic thickness; for further details see *Supplementary Information*.

| δ-layer species, depth | Physical thickness via SIMS (nm) | Electronic thickness via SX-ARPES (nm) |
|---|---|---|
| As, z = 2 nm | 2.01 ± 0.2 | 0.45 ± 0.04 |
| As, 3 nm | 2.22 ± 0.2 | 0.62 ± 0.10 |
| P, 2 nm | 2.24 ± 0.2 | 0.91 ± 0.21 |
| P, 3 nm | 2.86 ± 0.2 | 1.03 ± 0.35 |

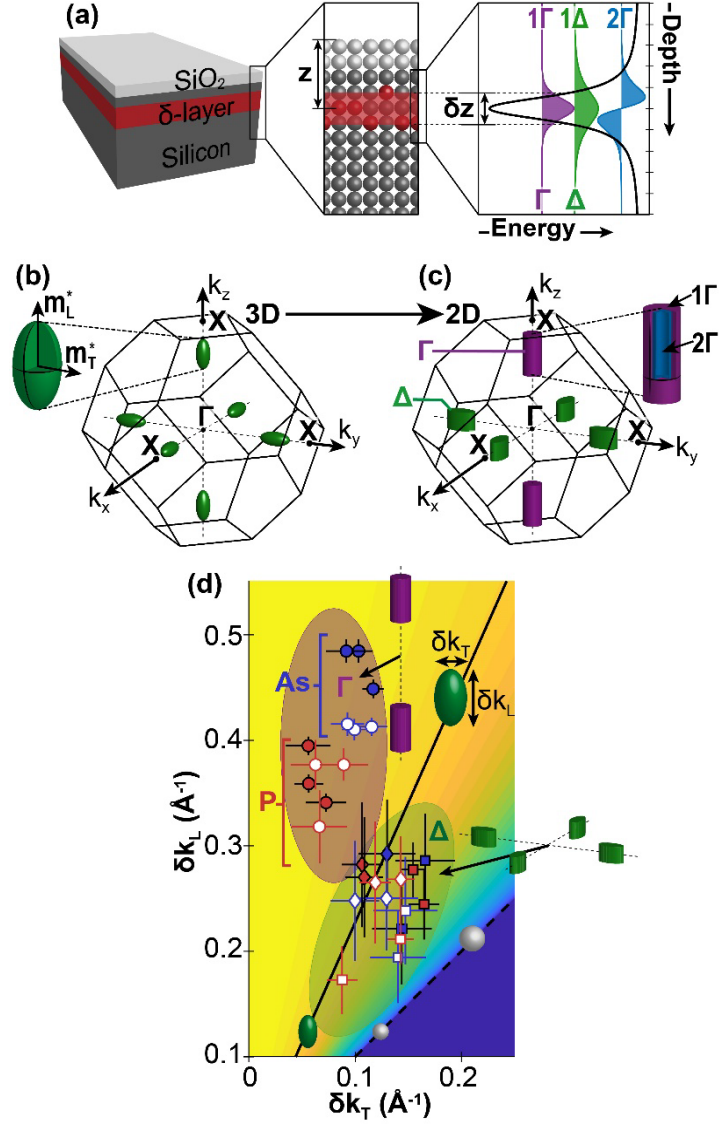

**Figure 1: Sample schematic and the evolution of the silicon conduction valleys vs δ-layer confinement.** (a) Schematic representation of our δ-layer samples, with a native oxide that forms due to ambient exposure. The silicon overgrowth thickness, $z$, is indicated, as is the electronic thickness of the δ-layer, $\delta z$. The δ-layer creates an approximately V-shaped potential well in the plane perpendicular to the δ-layer, which quantizes the out-of-plane and in-plane conduction valleys into a series of subbands denoted as $n\Gamma$ and $n\Delta$ respectively. The tic-marks on the depth axis indicate 1 nm steps. (b) Evolution of the silicon conduction valleys from 3D (6 degenerate, ellipsoidal valleys) to (c) 2D (4 Δ-valleys + 2 Γ-valleys). (d) Plot of the transverse ($\delta k_T$) versus longitudinal ($\delta k_L$) extent of the $k_x$ (diamonds, green region), $k_y$ (squares, green region) and $k_z$ valleys (circular markers, purple region). The filled and hollow markers represent data from 2 and 3 nm deep δ-layers respectively. The solid black line indicates the expected valley morphology for bulk, 3D silicon, whose gradient is equal to the mass anisotropy of silicon. The coloured background represents the eccentricity of the ellipsoid, which spans from zero (blue) to one (yellow).

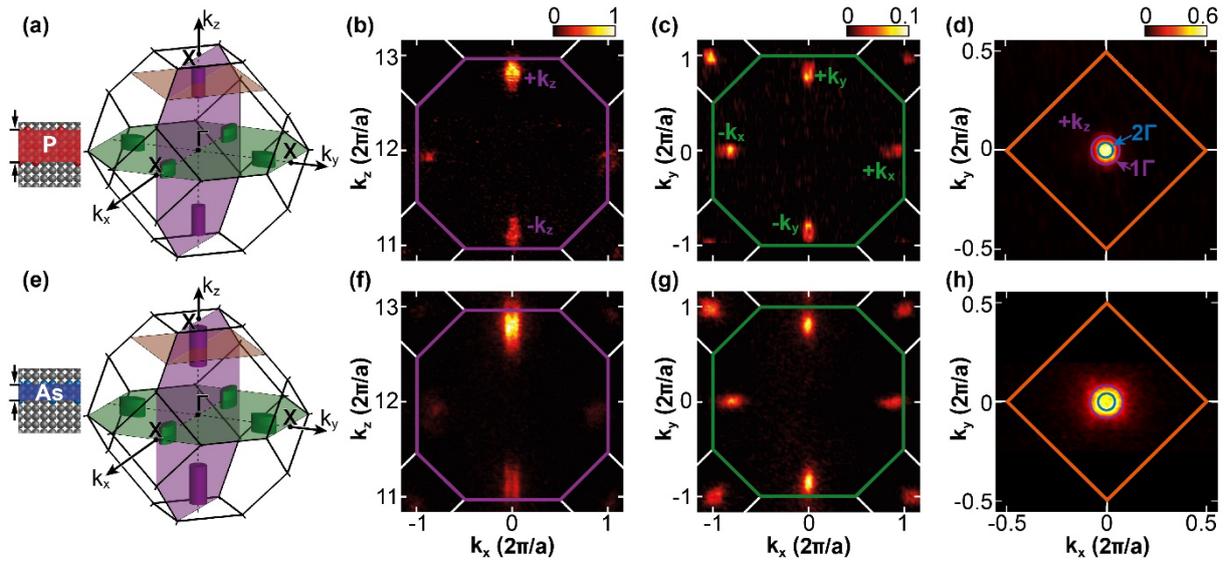

**Figure 2: Fermi-surface measurements of (upper row) phosphorous and (lower row) arsenic δ-layers with 2 nm silicon overgrowth.** (a,e) Schematic representations of the measured six conduction valleys of silicon embedded within the bulk fcc Brillouin zone, indicating 2D behaviour. Fermi surface slices for the phosphorous and arsenic δ-layer samples are shown along the following planes: (b,f) $k_x$-$k_z$ and (c,g) $k_x$-$k_y$ through the zone centre (Γ), (d,h) $k_x$-$k_y$ through the centre of the upper $k_z$ valley (see also the colour-coded slices on panels a,b). Fermi surfaces are integrated from -50 meV to $E_F$. In panel (d,h), the 1Γ and 2Γ states are denoted, based on the fits acquired in Figure 4.

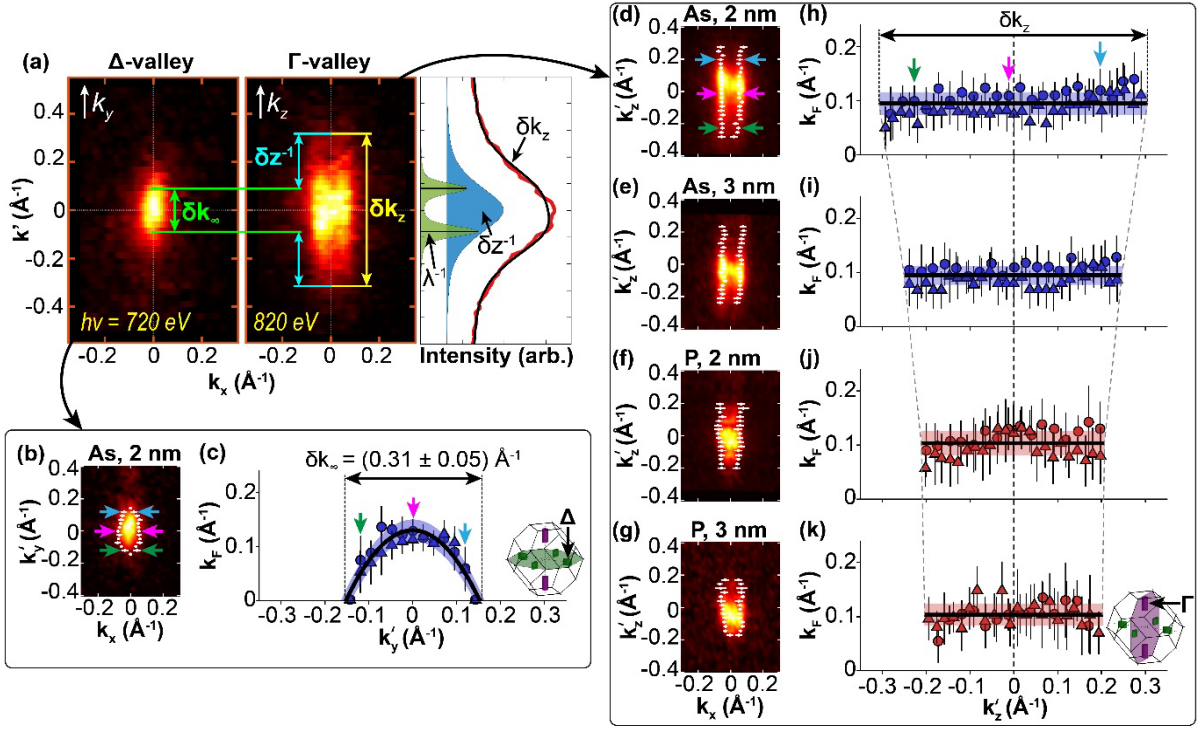

**Figure 3: Extracting δ-layer confinement from the longitudinal span of the Δ- and Γ-valleys.** (a) Visual representation of Equation 1, which is applied to extract the δ-layer confinement, $δz$, from the longitudinal extent of the Δ- and Γ-valleys. $δk_∞$ represents the longitudinal FWHM of the Δ-valley along the $k_y$-axis, which is broadened by the intrinsic mean free path (MFP), $λ$; $δk_z$ is the longitudinal FWHM of the Γ-valley along the $k_z$-axis, which includes both the MFP and confinement broadening. The line profile shows the Lorentzian deconvolution process (see *Supplementary Information* for more details), allowing the δ-layer confinement to be extracted from the $k_z$ response of the Γ-valley; a reasonable fit (black) to the data (red) can be achieved by convolving the green and blue contributions. (b) $k_x$-$k_y$ Fermi surface and (c) $k_F$ vs $k_y$ for the 2 nm arsenic δ-layer sample, where $k'_y = k_y - 0.94$ Å$^{-1}$. (d-g) $k_x$-$k_z$ Fermi surfaces ($hν = 350 – 410$ eV, integrated from -50 meV to $E_F$) for 2 and 3 nm phosphorous and arsenic δ-layers, as indicated. White dots indicate the cusps of the parabola in $k_x$ at each value of $k_z$, where $k'_z = k_z - 10.18$ Å$^{-1}$. (h-k) Plots of $k_F$ as a function of $k_z$ extracted from panels (a-d), whereas the triangle data points are taken from the same valley, but at a higher photon energy ($≈ 820$ eV). The inset of (c) and (k) indicates the valley that is probed; the green, pink, and blue arrows in panels (b,c) and (d,h) offer a guide to the eye. The best fit line is shown in black, with the shaded areas indicating the 1σ fit confidence.

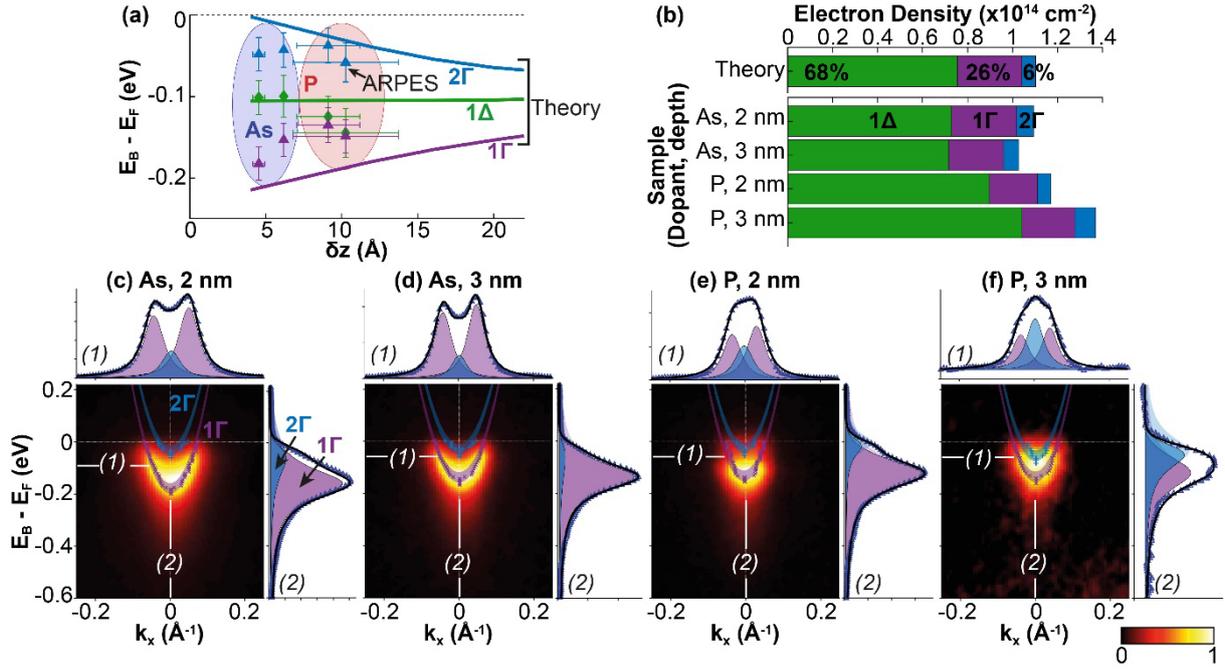

**Figure 4: Analysis of the conduction band quantisation for phosphorous and arsenic δ-layers.** (a) Measured and calculated energies of the δ-layer subbands. (b) Measured and calculated electron density for each δ-layer subband from the fits acquired in (c-f). For the theory, the percentages denote the occupation of each δ-layer subband; when the degeneracy is accounted for, each one of the six valleys have an equal occupancy of approximately 16.6%. (c-f) SX-ARPES measured Γ band dispersions for 2 and 3 nm phosphorous and arsenic δ-layers at $h\nu$ = 380 eV (which corresponds to the centroid of the Γ valley in Figure 3(a)). The purple and blue parabolas are the fits to the ARPES data, showing that the 1Γ and 2Γ states can be deconvolved. The line profile above each image shows the momentum distribution curves taken at (1); the line profile to the right, labelled (2), shows the corresponding fit to the energy distribution curve.

# Supplementary Information: Momentum space imaging of ultra-thin electron liquids in δ-doped silicon

Procopios Constantinou[*,1,2,5], Taylor Stock[1,9], Eleanor Crane[1,9], Alexander Kölker[1,9], Marcel van Loon[1,2], Juerong Li[3], Sarah Fearn[1,4], Henric Bornemann[1,2], Nicolò D'Anna[5], Andrew Fisher[1,2], Vladimir N. Strocov[5], Gabriel Aeppli[5,6,7,8], Neil Curson[1,9], Steven Schofield[†1,2]

[1] London Centre for Nanotechnology, University College London, WC1H 0AH, London, UK
[2] Department of Physics and Astronomy, University College London, WC1E 6BT, London, UK
[3] Advanced Technology Institute, University of Surrey, Guildford GU2 7XH, UK
[4] Department of Materials, Imperial College of London, London SW7 2AZ, UK
[5] Paul Scherrer Institute, 5232 Villigen, Switzerland
[6] Institute of Physics, Ecole Polytechnique Fédérale de Lausanne (EPFL), 1015 Lausanne
[7] Department of Physics, ETH Zürich, 8093 Zürich
[8] Quantum Center, Eidgenössische Technische Hochschule Zurich (ETHZ), 8093 Zurich, Switzerland
[9] Department of Electronic and Electrical Engineering, University College London, London WC1E 7JE, UK
[*] procopios.constantinou@psi.ch
[†] s.schofield@ucl.ac.uk

(Dated: Tuesday, 04 July 2023)

## Supplementary Information Contents

# (A) SIMS depth profiling of δ-layer samples

Time-of-flight secondary ion mass spectrometry (ToF-SIMS) measurements were made using an IONTOF ToF.SIMS5-Qtac100 system at the Surface Analysis Facility at Imperial College London. The base pressure was below $10^{-10}$ mbar, with a mass resolution $m/\Delta m =$ 10,000. The ToF-SIMS system was operated in static mode, with a 25 keV Bi$^+$ primary ion beam in high current bunch mode (HCBM). For depth profiling, a 250 eV, 20 nA Cs$^+$ sputter ion beam was used, with an impact angle at 45° to the sample surface, with an estimated sputter rate of $(0.03 \pm 0.02)$ nm/s. These settings were found to maximise the depth resolution and relative sensitivity factor, whilst minimising the atomic mixing and straggle depth [S1], [S2]. Depth profiles were made over a 300 μm × 300 μm sputter crater, where the analytical region was gated to the central 50 μm × 50 μm to minimise the artefacts of the non-uniform edges of the sputter crater. The depth of the SIMS craters was measured using a Zygo NewView 200 3D optical interferometer; this determines a uniform sputter rate for each SIMS measurement. To calibrate the secondary ion intensity to a volumetric density, a standard δ-layer sample with a known δ-layer density was used as reference.

Since the secondary ion yield for a given dopant species depends on the matrix it is embedded in, a normalisation procedure is required for a complete interpretation of the SIMS data. In the literature, accurate quantification has been studied for ultra-shallow dopant layers buried in a SiO$_2$ and Si stack, where either the SiO$_2$ or Si secondary ion signals are used for normalisation [S3]. Here, we normalise the dopant signal to the sum of the SiO$_2$ and Si signals, since the δ-layers lie within a convoluted SiO$_2$ and Si matrix.

Figure S1 shows the SIMS depth profiles of identically prepared phosphorous (red) and arsenic (blue) δ-layers buried 2 and 3 nm below the Si(001) surface. In comparing Figure S1(b,c) and Figure S1(d,e), we find that an arsenic δ-layer yields a better dopant confinement in the direction perpendicular to the δ-layer at comparative depths: (i) For the 2 nm deep arsenic and phosphorous δ-layer, a full width at half maximum (FWHM) of $(2.0 \pm 0.2)$ nm and $(2.2 \pm 0.2)$ nm is measured respectively; (ii) For the 3 nm deep arsenic and phosphorous δ-layer, a FWHM of $(2.2 \pm 0.2)$ nm and $(2.9 \pm 0.2)$ nm is measured, respectively. We attribute this to the smaller diffusion constant of arsenic in silicon [S4], [S5], which allows arsenic to remain more confined, relative to phosphorous, for equivalent thermal budgets of sample growth. It is important to add that the SIMS measure of the FWHM is an upper-bound estimate at best, as the depth profiles are inherently broadened dues to instrumental parameters, ion beam interactions and element-specific detections [S6], [S7]. Thus, we expect the actual δ-layers to be more confined than the results in Figure S1.

By integrating the area beneath the dopant signal of the SIMS depth profiles, the δ-layer sheet density can be determined. Here, the density is defined in terms of a monolayer (ML), which conveniently translates to the fractional coverage of dopants on the initial Si(001) surface, where $1\text{ ML} = 6.78 \times 10^{14}$ cm$^2$. For phosphorous and arsenic, we find a δ-layer density of $(0.36 \pm 0.04)$ ML and $(0.27 \pm 0.03)$ ML respectively; or $(2.5 \pm 0.3) \times 10^{14}$ cm$^{-2}$ and $(1.8 \pm 0.2) \times 10^{14}$ cm$^{-2}$, respectively. This difference in the saturation density is expected due to the differences in the dissociation chemistry of the dopant precursors [S8], [S9].

## (B) Hall measurements of δ-layer samples

Eight-terminal Hall bars were fabricated from the phosphorous and arsenic δ-layer samples after the SX-ARPES experiments. Figure S2(a) is a schematic of the final, etched Hall-bar mesa. After cleanroom processing, the samples were then electrically characterised at ≈ 10 K with Hall measurements, whose results are shown in Figure S2(b). By fitting to these data, we can quantify the carrier concentration and mobility for each sample. This is shown in Table S1. Note, we only show the data for the 3 nm deep arsenic δ-layer, as the 2 nm arsenic δ-layer sample did not display any transport due to open-circuit contacts; we attribute this to the potential failure of the etching process during cleanroom processing.

We find that the carrier concentration determined from the Hall experiments in Figure S2 is smaller than the δ-layer densities extracted from SIMS. This is expected, as not all incorporated donors provide electrons and there are several mechanisms for dose loss; deactivated donors in interstitials, vacancies or due to surface oxidation. Additionally, it is important to state that the samples were HF etched prior to the deposition of the Al contacts shown in Figure S2(a), which is required to achieve ohmic contact to the δ-layer. This inevitably brings the δ-layer closer to the surface and leads to another regrowth of the surface oxide; hence, a further dose loss of the δ-layers is expected relative to the initial SX-ARPES measurements. Nevertheless, the carrier concentrations remain well above the metal-insulator transition. In particular, since arsenic is known to suffer from clustering at high concentrations [S10], [S11], this further results in incomplete activation, and we attribute this to the smaller carrier concentration of the arsenic δ-layers compared to phosphorous in Table S1.

## (C) ARPES of samples with and without a δ-layer along XW

By aligning the ARPES measurement plane to coincide with the ΓX symmetry plane and tuning the photon energy to 380 eV, the $E_B(k_x)$ band dispersion along the XW symmetry line is mapped. Figure S3 shows a comparison of the ARPES data for a Si(001) sample without a δ-layer (Figure S3(a)), with a 2 and 3 nm deep arsenic (Figure S3(b)) and phosphorous (Figure S3(c)) δ-layer. The data immediately show that the sample without a δ-layer has no occupation of the conduction band. This is expected as the lack of any doping implies that the Fermi-level resides within the bandgap. However, both types of δ-layer samples show a bright conduction band pocket at the bulk X-point, which is a clear signature of the metallic nature of these δ-layers. A crucial point is that the photon energy is identical for all the ARPES data shown in Figure S3, so the inelastic mean free path (IMFP) of the emitted photoelectrons is constant. Thus, for a δ-layer buried deeper beneath the surface, we would expect it to yield a weaker spectral intensity, due to the sub-surface nature of the δ-layer. This is observed in Figure S3(b) and Figure S3(c), where the 3 nm deep δ-layers both show a diminished intensity relative to the 2 nm deep δ-layer. This provides a clear indication that the occupied conduction band state originates from the sub-surface δ-layer.

## (D) Quantifying the δ-layer carrier density

We discuss here the procedure used to quantify the carrier concentration from the ARPES Fermi surface. All the relevant ARPES data are shown in Figure S4 for both the arsenic and phosphorous δ-layer samples, where $k_x$-$k_y$ slices are taken through the conduction valleys. Note, in the bottom panel of Figure S4(d), the $+k_x$ valley does not appear as expected and we attribute this to a slight rotational misalignment during the acquisition of the ARPES data.

The total area enclosed by each valley is determined consistently by thresholding the ARPES data at two iso-contour limits; values at which the maximum intensity drops to 25% (upper bound estimate of the enclosed area) and 50% (lower bound estimate of the enclosed area). These two limits are shown in Figure S4(a-d) as shaded regions around each valley; coloured purple for the $+k_z$ valley (middle panel) and green for the in-plane ($\pm k_x, \pm k_y$) valleys (bottom panel). An estimate of the carrier concentration, $n_e$, is then determined by applying *Luttinger's Theorem* [S12], which states that the total area enclosed by the Fermi surface is directly proportional to the number density of electrons. A simple expression for this can be derived from the free-electron (Sommerfeld) model, where:

$$n_e = 2 \frac{1}{(2\pi)^j} V_{kj}. \qquad (5)$$

Here, $V_{kj}$ is the $j$-dimensional volume of $k$-space, where in the case for the two-dimensional δ-layer system, $j = 2$. The result of applying Equation (5) is shown in the upper-panel of Figure S4(a-d), which shows a breakdown of the carrier concentration for each one of the six valleys. We find that within the experimental uncertainty, there is an equal occupancy amongst each valley, where the sum yields the total carrier concentration of each δ-layer sample. Since the $n = 2$ state of the Γ-valley cannot be clearly resolved within the Fermi-surfaces shown in the middle panel of Figure S4(a-d), the contribution of the 2Γ state is extracted from the fits to Figure S6(a-d). A free-electron like (or circular) Fermi-surface is assumed. In doing so, the carrier concentration estimate slightly increases by 10% and the total carrier concentration for all δ-layer samples is found to lie within $(0.88 \pm 0.10) \times 10^{14}$ cm$^{-2}$.

Alternatively, it is known that a more accurate method to determine the total area enclosed by the Fermi surface is to extract the maximum gradient of the energy integrated photoemission intensity from the experimental ARPES data [S13]. This so-called gradient method is based on the theoretical many-body definition of the Fermi surface and this analysis was implemented for the 2 nm arsenic δ-layer sample, using the data in Figure S4(a). In doing so, the total carrier concentration was found to be approximately the same; ~15% higher.

## (E) Quantifying the δ-layer thickness by fitting the k$_z$ response

Here, we compare the two methods of quantifying the δ-layer confinement ($\delta z$) discussed in the main manuscript and show that they are in mutual agreement (Figure S5(g)). The first method is to apply Equation (1) from the main manuscript, where $\delta z$ is the reciprocal of the difference in the longitudinal span of the Γ- and Δ-valleys (Figure 3). The second method relies on fitting the $k_z$ response of the data to the convolution of a pair of mean free path (MFP) broadened Lorentzian's, to another Lorentzian whose width is proportional to $\delta z^{-1}$ (Figure

S5(a)). A Lorentzian curve shape is used since the Fourier Transform of an exponentially decaying wavefunction is Lorentzian, and a pair of curves due to the $\pm k_z$ momenta of the initial states that are in superposition and form a standing wave perpendicular to the δ-layer. The TPP-2M formalism for silicon [S14] is used to estimate the photoelectron MFP, $\lambda_e$, via the incident photon energy, $h\nu$; we find $\lambda_e$ spans 1.2 – 2.0 nm for $h\nu$ between 380 – 820 eV. Furthermore, we neglect the contribution from the spectrometer energy resolution, which translates into a momentum broadening of $\Delta k \approx 0.02 \text{Å}^{-1}$ about the same $h\nu$ range. The initial position of the MFP-broadened peaks is estimated by finding the corresponding Fermi wave-vector, $k_F$, of the Δ-valleys in $k_x$-$k_y$. A least-squares approach is employed to find the optimal value of $\delta z$ required to fit the $k_z$ response of both the Γ- and Δ-valleys.

Horizontal ($k_z$) and vertical ($k_x$) cuts of the Fermi-surface are shown in Figure S5(a), with the 1Γ and 2Γ states being identified. We observe that the first quantum well state (1Γ) brackets the second quantum well state (2Γ), and that the intensity of the $k_z$ line-profile across the Γ-valley is dominated by the 1Γ contribution. Thus, when fitting the $k_z$-response of the data, we found near identical results for $k_x$ intervals that are either localised around 1Γ, or integrated across the whole Γ-valley. Additionally, by taking the Fourier Transform of the probability density function solutions from the Schrödinger-Poisson model of δ-layers derived in Figure S7 (discussed further in Section G), we find that they also provide a good agreement to the $k_z$ ARPES spectral response. This is identically shown for the Δ-valley in Figure S5(b), where the 1Δ singlet state can be identified. Furthermore, we again find that the Fourier Transform coincides with theoretical expectations, which also appears more localised along $k_z$; the sharper $k_z$-response is associated with the 1Δ wave function being broader in real-space (see Figure S7).

Figure S5(c-f) shows the best estimate of $\delta z$ using the model fit approach outlined in Figure S5(a-b) and discussed above. We find that arsenic δ-layers yield a better confinement relative to phosphorous and that the fits to both the Γ- and Δ-valleys yield consistent values of $\delta z$. Furthermore, Figure S5(g) confirms that either of the approaches discussed above can be used to estimate $\delta z$, as the values are in mutual agreement. This works because the convolution of Lorentzian's yields a Lorentzian whose width is the arithmetic sum of the component widths.

## (F) Γ- and Δ-band fitting procedures

The procedures used to fit the ARPES data shown in Figure 4 of the main manuscript are discussed here. A summary of the Γ-band fits are shown in Figure S6(a-d) and the Δ-band fits are shown in Figure S6(e-h), for all δ-layer samples. Each panel shows the ARPES data, overlaid with the best fit parabolic bands, where the number in brackets identifies the type of cut that is taken through the ARPES data. There are two cuts used in the fitting of the ARPES data; (i) momentum distribution curves (MDCs) or horizontal cuts (shown above the ARPES data) and (ii) energy distribution curves (EDCs) or vertical cuts (shown to the right of the ARPES data). We find that the Γ-band fits can be deconvolved into two discrete states, denoted as 1Γ and 2Γ, whereas the Δ-band fits are deconvolved into a single state, 1Δ.

Initially when trying to fit the Γ-bands consistently with a single state, the middle regions of the MDCs in Figure S6(a-d) (labelled as (1)-(3)) were found to be significantly underestimated, with no satisfactory fit being achieved. It was also found that when using pure Gaussian functions, the MDC cuts would not fit well due to the extended tails that are formed by each of the peaks. The best fits made for each sample are shown in Figure S6(a-d), where the fits are comprised of multiple peaks that follow a parabolic path in $k_x$, derived from two discrete sub-bands, which we denote as 1Γ and 2Γ. A least-square fitting algorithm is employed in MATLAB that consists of fitting a series of Voigt functions, equally weighted in terms of its Gaussian and Lorentzian parts, that each follow an initially parabolic path in $E_B(k_x)$ until a satisfactory fit converges through each MDC cut. This form was chosen as it combined the experimental resolution and intrinsic broadening effects, whilst optimally fitting the data. The fitting algorithm starts at an MDC cut taken at $-0.2$ eV, and progressively moves upwards through the Γ-state fitting up to four Voigt functions (for $n = 2$ states) until it hits the Fermi-level, after which it terminates. The results of this are overlaid on the ARPES images of Figure S6(a-d). To ensure consistency, all fits were performed using similar constraints. The FWHM is constrained to $(0.06 \pm 0.03)$Å$^{-1}$ and the best fit parabola for all δ-layer samples yielded an effective mass of $(0.12 \pm 0.03)m_e$, which lies close to the expected value of $0.19m_e$ [S15]. Furthermore, the EDC cut through the centre of mass of the Γ-band in Figure S6(a-d) (labelled as (4)) also shows a satisfactory fit when using two components. It is important to note that the model curve used to fit the EDC case is more complicated; it is the convolution of a Voigt function with a Gaussian broadened Fermi-Dirac distribution (the Gaussian broadening is a consequence of the 60 meV energy resolution). Therefore, the EDC curve fits appear asymmetrical and drop to zero near $E_B = E_F$.

For the Δ-band fits shown in Figure S6(e-h), the fits are simpler as a single state was found to yield satisfactory results. However, due to the higher effective mass of the Δ-band, its much smaller curvature makes it more difficult to track the peak locations, which is why the best fit parabola has a larger uncertainty, relative to the Γ-band fits. Because of this, the fits were made to each EDC cut from left to right, using the same algorithm discussed above. For the arsenic δ-layer samples in Figure S6(e-f), the peak position appears to be constant when fitting the EDCs, however, this is due to the base of the Δ-band being very near the Fermi-edge and its very low curvature at the base of the parabola makes it seem flat. For phosphorous δ-layer samples in Figure S6(g-h), the parabolic minimum is further away from the Fermi-edge and so the slight curvature of the Δ-band becomes apparent. To ensure consistency, all fits were performed using similar constraints. The FWHM of the EDC fits were constrained to $(0.16 \pm 0.04)$ eV and the best fit parabola for all δ-layer samples yielded an effective mass of $(0.90 \pm 0.05)m_e$, which lies close to the expected value of $0.98m_e$ [S15].

## (G) Self-consistent Schrödinger-Poisson model of δ-layers

The theoretical model being compared to the experimental data in Figure 4 of the main manuscript is outlined here. The model employs an effective mass approximation [S16] and accounts for dopant segregation in realistic δ-layers [S17]. Given the latter, the model also neglects the inter-valley coupling, which becomes vanishingly small for δ-layers wider than 0.2 nm [S17].

In highly doped δ-layers, a strong electrostatic potential is induced that confines electrons to the plane. The high concentration of donor electrons leads to a substantial overlap of their wavefunction and screening of the attractive cores. As a result, the local disorder is smoothed out. It is therefore convenient to ignore the placement of individual atoms and describe the layer by a constant average charge density. The system now exhibits planar symmetries, where the electrostatic potential only varies in the plane-perpendicular direction. The infinite δ-layer is thus described by one-dimensional Schrödinger-like equations:

$$\left(-\frac{\hbar^2}{2m_\parallel}\frac{d^2}{dz^2} + V(z)\right)F_\Gamma(z) = \epsilon_\Gamma F_\Gamma(z), \tag{7}$$

$$\left(-\frac{\hbar^2}{2m_\perp}\frac{d^2}{dz^2} + V(z)\right)F_\Delta(z) = \epsilon_\Delta F_\Delta(z). \tag{8}$$

Here, $m_\parallel \approx 0.98 m_e$ corresponds to the longitudinal anisotropic electron mass in silicon and $m_\perp \approx 0.19 m_e$ the transverse effective mass given in terms of the bare electron mass $m_e$. Note that the equations are not given in terms of the full electronic wavefunction $\psi(\mathbf{r}) = F(\mathbf{r})\psi(\mathbf{k},\mathbf{r})$, but instead of envelope functions $F_i$. The complete solutions of the wave equations include rapid Bloch oscillations that can be omitted in the effective mass approximation. The Schrödinger-like equations are re-written as systems of first order linear equations. The eigenvalues $\epsilon_i$ and envelope functions are computed using a root-finding Schrödinger solver and the second order Runge-Kutta method. Note that the model neglects inter-valley coupling. This is appropriate for the current samples, as the valley splitting of the 1Γ band becomes vanishingly small for δ-layers wider than 0.2 nm [S17].

Assuming the electron wavefunctions are separable, the plane-parallel solutions to the wave equation correspond to a two-dimensional electron gas (2DEG). Each of the quantised bands from the plane-perpendicular envelope functions can thus be occupied by electrons obeying the state filling rules of a 2DEG. Taking the two- and four-fold degeneracy of the bands into account, the following set of equations are obtained:

$$\begin{aligned}2(\epsilon_F - \epsilon_{n\Gamma})D_\Gamma(\epsilon) &= N_{n\Gamma}, \\ 4(\epsilon_F - \epsilon_{n\Delta})D_\Delta(\epsilon) &= N_{n\Delta},\end{aligned} \tag{9}$$

where $D_\Gamma(\epsilon) = m_\perp/\pi\hbar^2$ and $D_\Delta(\epsilon) = \sqrt{m_\parallel m_\perp}/\pi\hbar^2$ are the density of states per unit area of the corresponding bands. The Fermi energy is found by the charge neutrality condition, i.e. the total number of electrons $N_e = \sum_i N_i$ equals the total number of donors $N$, and by finding a self-consistent solution for:

$$\epsilon_F = \frac{N + 2D_\Gamma \sum_j^{M_\Gamma} \epsilon_j^\Gamma + 4D_\Delta \sum_k^{M_\Delta} \epsilon_k^\Delta}{2M_\Gamma D_\Gamma + 4M_\Delta D_\Delta}, \tag{10}$$

where $M_i$ are the number of bands below the Fermi energy. The electrons are distributed to the different bands according to Equation (9).

The potential $V(z) = V_H[n(z)] + V_{XC}[n(z)]$ is obtained from the resulting electron carrier distribution. The exchange-correlation potential $V_{XC}$ is computed with the XC functional by *Ceperley and Adler* and by employing scaled atomic units in the parametrization by *Perdew and Zunger* [S18]. The Hartree potential $V_H$ is obtained from Poisson's equation by applying the finite difference method. The method maps the problem onto a grid, such that the second derivative of the potential becomes:

$$(\partial_z^2 V)_i \approx \frac{V_{i-1} - 2V_i + V_{i+1}}{dz^2}, \tag{11}$$

where $i$ are the indices of the grid. By adopting the matrix representation and imposing the boundary conditions $\lim_{z \to \pm\infty} V(z) = 0$, the Poisson equation takes the form:

$$\frac{1}{dz^2} \underbrace{\begin{pmatrix} -2 & 1 & 0 & \cdots & 0 \\ 1 & -2 & 1 & \cdots & 0 \\ 0 & \ddots & \ddots & \ddots & \vdots \\ \vdots & \ddots & 1 & -2 & 1 \\ 0 & \cdots & 0 & 1 & -2 \end{pmatrix}}_{=M} \cdot \begin{pmatrix} V_0 \\ V_1 \\ \vdots \\ V_{b_z-1} \\ V_{b_z} \end{pmatrix} = \frac{e}{\epsilon_r \epsilon_0} \begin{pmatrix} n_0 \\ n_1 \\ \vdots \\ n_{b_z-1} \\ n_{b_z} \end{pmatrix}, \tag{12}$$

where $b_z$ is the number of discrete points centred around $z = 0$. Multiplication with the inverse matrix $M^{-1}$ yields the Hartree potential.

Unlike the effective mass approximation in Ref. [S16], the donor atoms are not confined to one doping layer but are subject to a sharp Gaussian segregation profile, in line with Ref. [S17]. Adding up the opposing electrostatic potentials from dopants and electrons and the exchange-correlation contribution gives the total potential of the system $\tilde{V}(z)$. The calculation is repeated with $V(z) = \tilde{V}(z)$ until self-consistency is reached. It occurs when $\int [V_{\alpha+1}(z) - V_\alpha(z)]dz \approx 0$ is satisfied, where $\alpha$ and $\alpha + 1$ refer to successive iterations.

Figure S7 summarises the results obtained when using the self-consistent Schrödinger-Poisson model of δ-layers. The wave-functions of the occupied Γ- and Δ-band states are shown in Figure S7(a-b), along with their corresponding carrier concentrations. We find an equal occupancy amongst the Γ- and Δ-bands for metallic δ-layers, whose density is of the order $10^{14} \text{cm}^{-2}$. Figure S7(c) also shows the evolution of the energy minima of each sub-band as a function of the FWHM of the δ-layer; as the δ-layer becomes broader, the effect of energy quantisation becomes smaller.

# Supplementary Information References

# Supplementary Information Figures

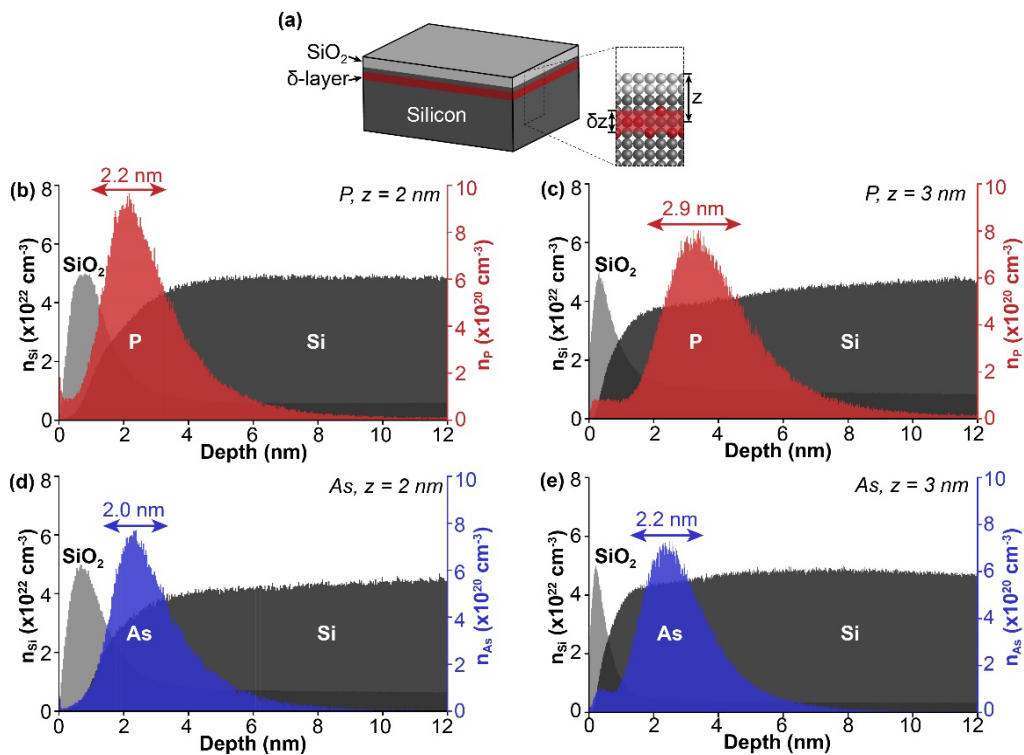

**Figure S1: SIMS depth profiles of ultra-shallow δ-layer samples.** (a) Schematic representation of the samples prepared; $z$ is the silicon overgrowth thickness, noting surface oxidation will have occurred, and $\delta z$ is the thickness of the δ-layer. (b,c) Secondary ion mass spectrometry (SIMS) of phosphorous δ-layer samples fabricated with (b) 2 nm, (c) 3 nm of silicon overgrowth. (d,e) SIMS of the arsenic δ-layer samples fabricated with (d) 2 nm, (e) 3 nm of silicon overgrowth. The Si(dark grey), SiO$_2$ (light grey), P (red) and As (blue) signals are shown. The Si signal is derived from the Si$_3$ fragmentation species. The data are calibrated to known fluence standards.

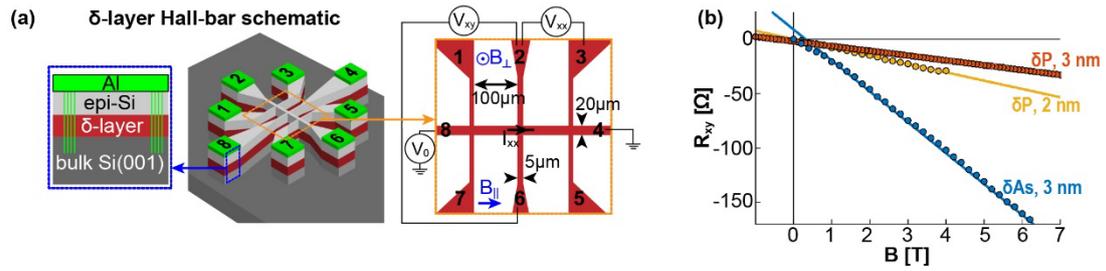

**Figure S2: Summary of the magnetotransport measurements of δ-layer samples.** (a) Schematic of the Hall-bar mesa created after cleanroom processing. A cross-section of the δ-layer device stack is shown to the left, contacted with aluminium vias. A zoom in of the Hall-bar geometry is shown to the right, where the orientation of the magnetic field is shown in blue, and the measurement terminals for the current and voltages $I_{xx}$, $V_{xx}$ and $V_{xy}$ are indicated. (b) Hall measurements with linear fits, allowing the carrier concentration to be determined. All measurements were made at ≈10 K.

**Table S1: Electrical transport characteristics extracted from the data shown in Figure S2.** The carrier concentration ($n$) and mobility ($\mu$) are determined from the linear fits shown in Figure S2(b).

| δ-layer species, depth | Carrier concentration ($n$) ($10^{14}$/cm$^2$) | Mobility ($\mu$) (cm$^2$ / V s) |
|---|---|---|
| As, z = 2 nm | N/A | N/A |
| As, 3 nm | 0.22 ± 0.01 | 77 ± 2 |
| P, 2 nm | 0.82 ± 0.02 | 62.53 ± 0.04 |
| P, 3 nm | 1.42 ± 0.01 | 31.60 ± 0.01 |

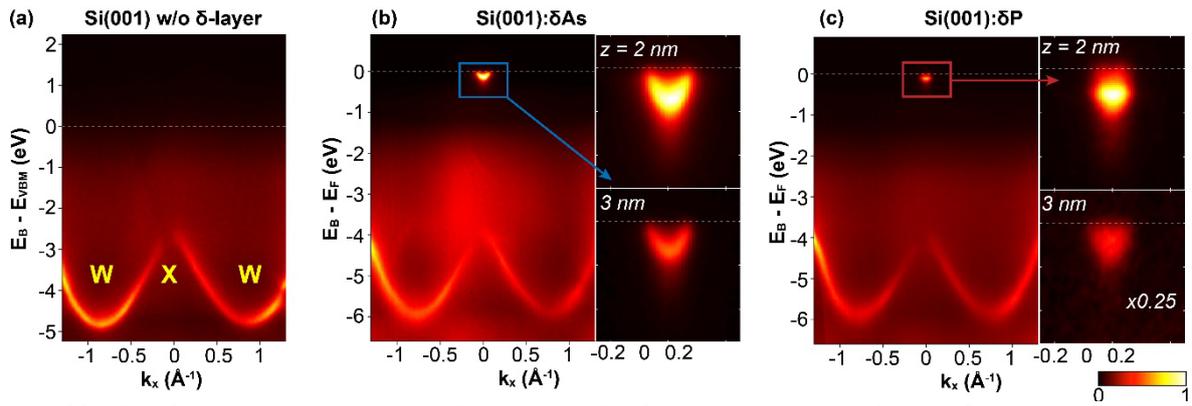

**Figure S3: ARPES measurements along the XW symmetry line for samples with and without a δ-layer.** (a) ARPES data for a Si(001) sample without a δ-layer, showing that there are no occupied conduction band states. The binding energy origin has been set to the valence band maximum. (b)-(c) ARPES data for a 2 nm deep (b) arsenic and (c) phosphorous δ-layer buried in Si(001). The binding energy origin is set to the Fermi-edge, where a zoom-in of the occupied conduction band is shown at two different depths. The colour scale is consistent across the figure and the data is take using a photon energy $h\nu$ = 380 eV.

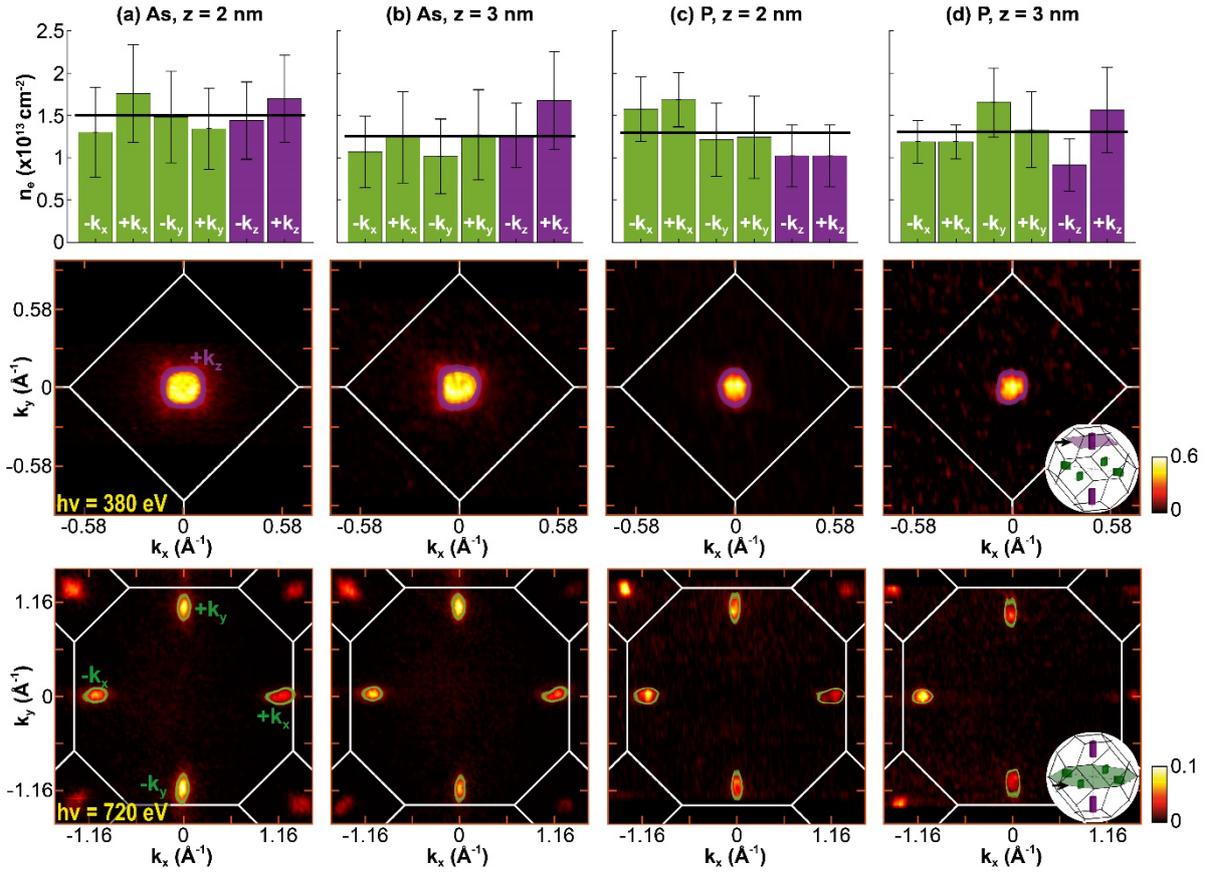

**Figure S4: Summary of the SX-ARPES determined carrier concentration for each $\delta$-layer sample.** The dopant species and depth for each sample is labelled and ordered such that each column represents a summary of the data obtained; (a-b) arsenic and (c-d) phosphorous $\delta$-layers buried (a,c) 2 nm and (b,d) 3 nm deep. (top) Breakdown of the carrier concentration for each of the six valleys, where the solid black line shows the average occupancy of each valley. (middle) ARPES iso-energetic slice extracted through $k_x$-$k_y$ for the $+k_z$ valley at $h\nu$ = 380 eV (bottom) ARPES iso-energetic slice extracted through $k_x$-$k_y$ for the in-plane ($\pm k_x, \pm k_y$) valleys. The iso-slices are integrated over the binding energy, $E_B$, from -250 meV to $E_F$. The iso-contours are also shown, which show the best estimate for the total enclosed area for each valley.

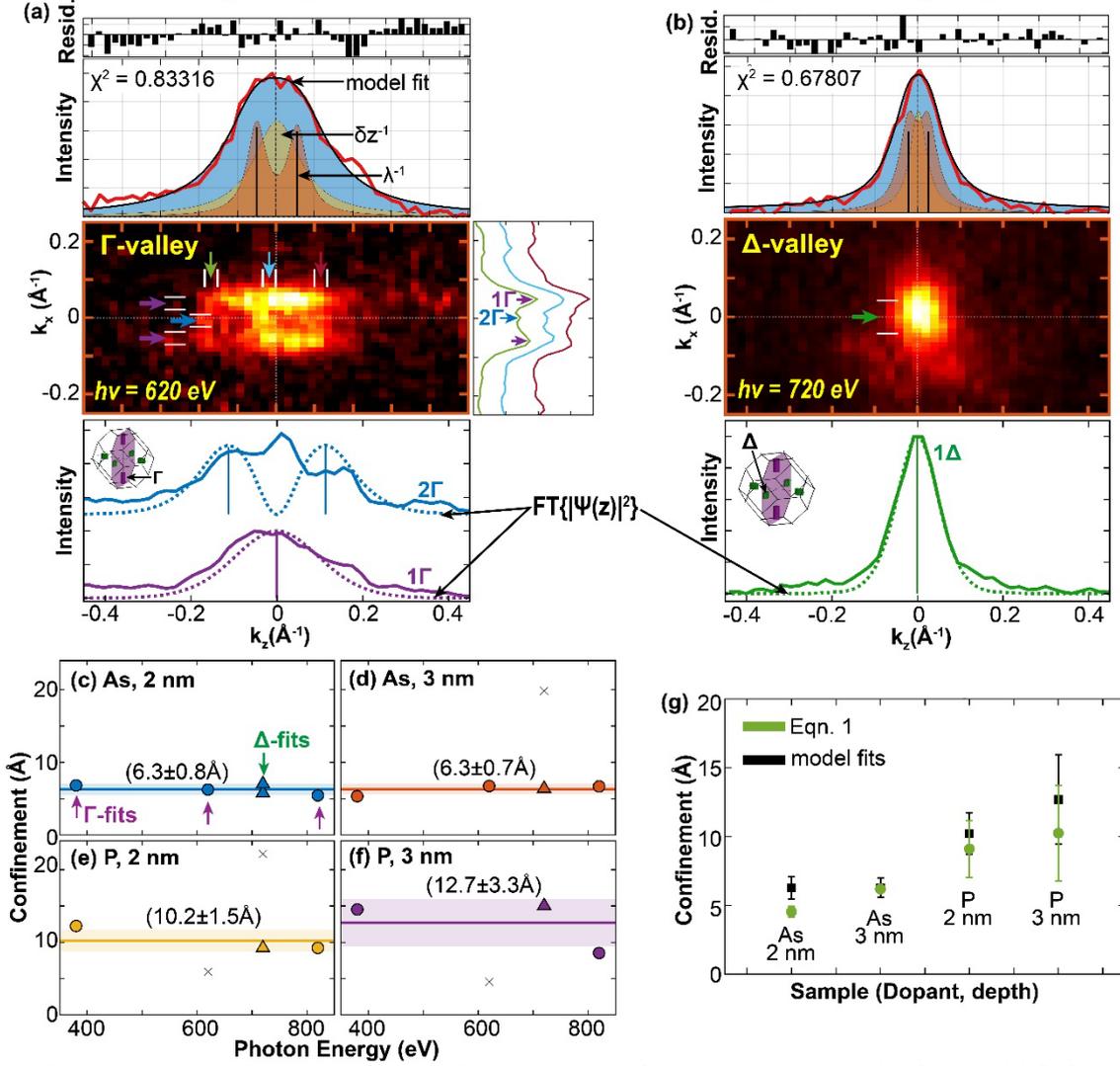

**Figure S5: Extracting the δ-layer confinement from the ARPES $k_z$-response.** (a) Example of the model fit for one Γ-valley dataset, whose chi-squared ($\chi^2$) value is stated: (middle) $k_x$-$k_z$ Fermi-surface; (upper) the $k_x$-integrated line-profile which shows the best fit (blue) to the data (red) by convolving the MFP-broadened (orange) and confinement-broadened (yellow) Lorentzian's; (lower) $k_z$ and (right) $k_x$ line profile cuts taken at three different points and colour-coded by the arrows. The white lines indicate the $k$-integration windows for the horizontal and vertical line-cuts. In the lower $k_z$ line profiles, the Fourier Transform (FT) of the probability density function solutions from a Schrödinger-Poisson model of δ-layers (Figure S7) is in good agreement to the data. (b) The same as (a), but for a Δ-valley dataset. The ARPES $k_z$-response appears sharper since the 1Δ wave function is broader in real-space (see Figure S7). (c-f) Plot of the δ-layer confinement vs photon energy for the model fits defined in (a) and (b) for all samples. The circle and triangle data-points are from the Γ- and Δ-fits respectively. The black crosses indicate outlier fits; the $k_z$ line profiles were noisy or had very poor fits to the data here. The shaded areas represent the uncertainty from the fitted range of values. (g) Comparison of the δ-layer confinement found using Equation 1 in the main manuscript (green) and the model fits (black) shown in (c-f). Both methods provide estimates to the δ-layer confinement that are in mutual agreement.

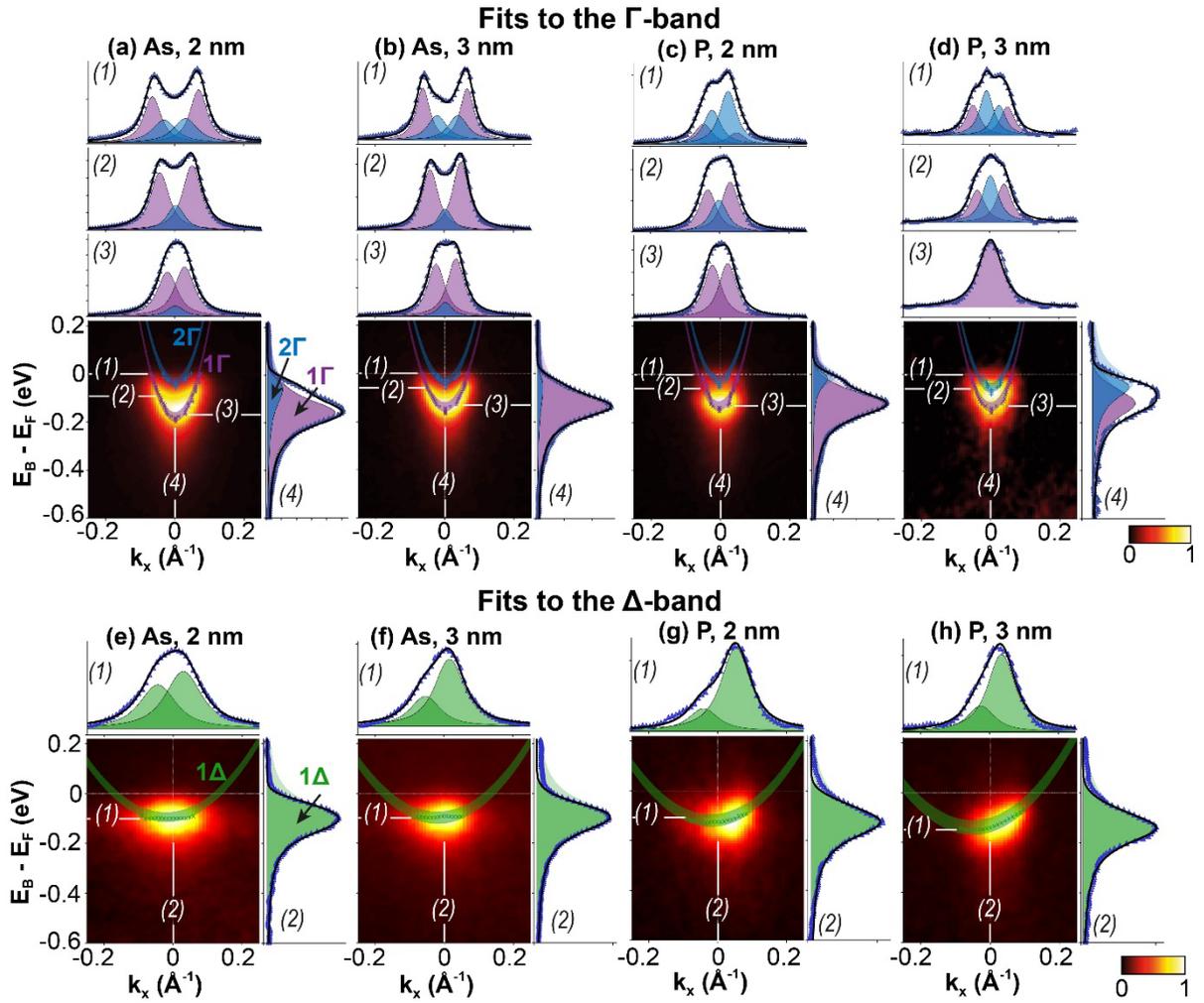

**Figure S6: Summary of the ARPES curve fits to the (upper) Γ- and (lower) Δ-bands of the 2 and 3 nm arsenic and phosphorous δ-layer samples.** The dopant species and depth for each sample is labelled and ordered such that each column represents a summary of the data and fits for each δ-layer sample. (a-d) Summary of the fits to the Γ-bands along the bulk XW symmetry line at a photon energy $h\nu$ = 380 eV. The purple and blue parabolas are the fits to the ARPES data, showing that the 1Γ and 2Γ states can be deconvolved. The line profiles above the ARPES image show the momentum distribution curves (MDCs) taken at three different binding energies, labelled (1)-(3); the line profile to the right, labelled (4) shows the corresponding fit to the energy distribution curve (EDC), taken through the centre of mass of the Γ-band. (e-h) Summary of the fits to the Δ-bands along the bulk ΓX symmetry line at a photon energy $h\nu$ = 720 eV. The green parabola shows the best fit to the 1Δ-band. The line profile above the ARPES image shows the fitted MDC, labelled (1) and the line profile to the right, labelled (2), shows a similarly fitted EDC, both of which are taken through the centre of mass of the Δ-band.

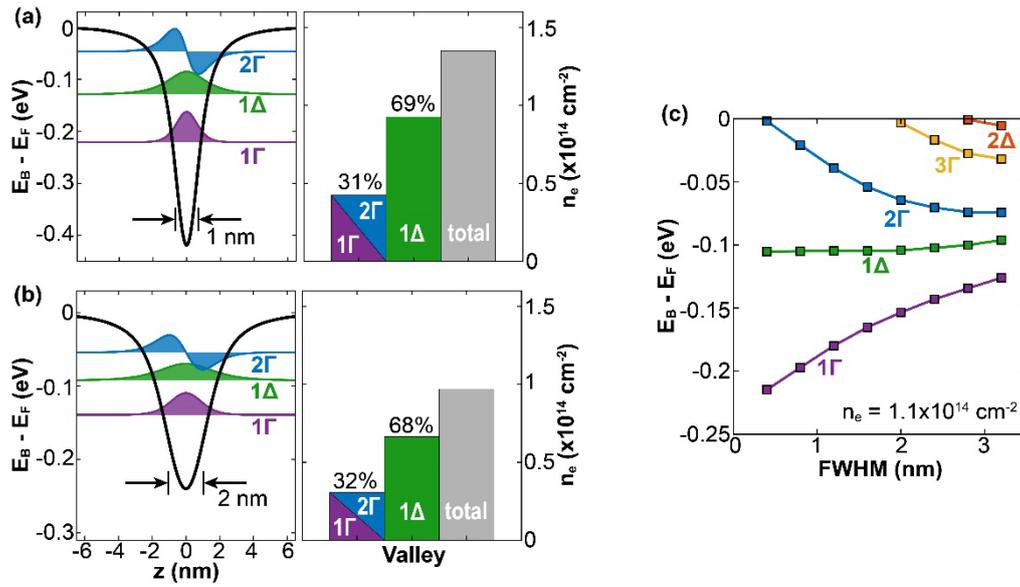

**Figure S7: Summary of the solutions using a self-consistent Schrödinger-Poisson model of δ-layers.** (a-b) Plot of the (left) wave-function and (right) carrier concentration solutions when using a charge distribution that has a FWHM of (a) 1 nm and (b) 2 nm. The occupancy of the Γ- and Δ-bands are shown as percentages, which show an equal occupancy (≈ 16%) for each sub-band. (c) Plot of the energy minima of each sub-band as a function of the FWHM of the charge distribution, using a δ-layer with a layer density of $1.1 \times 10^{14} cm^{-2}$.

S17